# Observation of solar high energy gamma and X-ray emission and solar energetic particles

**A Struminsky[1] and W Gan[2]**

[1] Space Research Institute, RAS, Profsoyuznaya st., 84/32, Moscow 117927, Russia
[2] Purple Mountain Obervatory, CAS, 2 West Beijing Road, Nanjing 210008, China

E-mail: astrum@iki.rssi.ru

**Abstract**. We considered 18 solar flares observed between June 2010 and July 2012, in which high energy >100 MeV γ-emission was registered by the Large Area Telescope (LAT) aboard FermiGRO. We examined for these γ-events soft X-ray observations by GOES, hard X-ray observations by the Anti-Coincidence Shield of the SPectrometer aboard INTEGRAL (ACS SPI) and the Gamma-Ray burst Monitor (GBM) aboard FermiGRO. Hard X-ray and $\pi^0$-decay γ-ray emissions are used as tracers of electron and proton acceleration, respectively. Bursts of hard X-ray were observed by ACS SPI during impulsive phase of 13 events. Bursts of hard X-ray >100 keV were not found during time intervals, when prolonged hard γ-emission was registered by LAT/FermiGRO. Those events showing prolonged high-energy gamma-ray emission not accompanied by >100 keV hard X-ray emission are interpreted as an indication of either different acceleration processes for protons and electrons or as the presence of a proton population accelerated during the impulsive phase of the flare and subsequently trapped by some magnetic structure. In-situ energetic particle measurements by GOES and STEREO (High Energy Telescope, HET) shows that five of these γ-events were not accompanied by SEP events at 1 AU, even when multi-point measurements including STEREO are taken into account. Therefore accelerated protons are not always released into the heliosphere. A longer delay between the maximum temperature and the maximum emission measure characterises flares with prolonged high energy γ-emission and solar proton events.

## 1. Introduction

At present two competitive paradigms of solar energetic particle (SEP) origin exist – acceleration directly in the flare site or by shock wave driven by a coronal mass ejection (CME) [1]. Active discussions on a relative role of flares and CME's for SEP acceleration and propagation are continuous till now. Additional arguments supporting one or another paradigm can be obtained investigating new events and using new instruments earlier unavailable.

More than 20 years ago the prolonged high energy γ-emission was discovered in solar flares of June 1991 by using sensitive detectors aboard the ComptonGRO and GAMMA-1 spacecraft designed for astrophysical studies [2-6]. The registration of $\pi^0$ decay γ-emission during several hours after the flare proves a possibility of prolonged proton interaction at the Sun and is an additional argument in favor of the flare origin of SEP. The launch of the Fermi Gamma-Ray Observatory (FermiGRO) into the orbit, which is the new generation of gamma-observatories, in 2008 allowed hoping that new observations of prolonged high energy γ-emission would be performed in the 24th solar cycle.

The Large Area Telescope (LAT) aboard FermiGRO observed high energy >100 MeV solar γ-emission in 18 flares [7]. For the first time such emission was attributed to solar flares of X-ray classes of C and M. The record duration, more than 20 hours, of the solar >100 MeV γ-emission was registered on March 7, 2012[8]. The $\pi^0$ decay is the most possible mechanism of its origin [7, 8]. Such a prolonged solar γ-emission raises three questions: "Are protons producing $\pi^0$ accelerated in flare impulsive phase and then trapped? Are they accelerated in several episodes during decay phase? Are they a possible source of solar protons propagating in the heliosphere?" In order to answer first two questions we need to check solar observations in SXR and HXR wave bands. Light curves of SXR emission may show episodes of additional energy release (plasma heating). Bursts of HXR emission will show episodes of electron acceleration and roughly maximal energy of electron acceleration. Note, that in one loop evaporation model of solar flares maxima of time derivatives of SXR intensity correspond to maxima of HXR intensity (the Neupert effect [9]). An answer to the third question would be positive, if solar high energy γ-emission is observed during solar proton events.

In this work we consider the 18 FermiGRO solar γ-events [7] in soft X-rays (SXR) by data of GOES spacecraft and hard X-rays (HXR) by ACS SPI and GBM data (table 1) as well as an association of these events with solar proton events observed in the heliosphere (table 2) to answer three questions stated above. The GBM registers solar HXR and gamma emission from few keV to about 1 MeV by NaI scintillators and from ~150 keV to ~30 MeV by BGO crystals. The ACS SPI is a BGO crystal detector and registers primary (solar) and secondary (caused by Solar Energetic Particles (SEP)) hard X-rays >100 keV. Methodological problems of ACS SPI data application for studies of solar flares were considered recently in [10]. The INTEGRAL spacecraft is in an orbit with perigee of 10000 km and apogee of 153000 km and has an orbit period of 72 hours, it is out of the radiation belts most of the time.

Data of ACS SPI may show processes of energy release and interaction of accelerated particles, which are not observed by the FermiGRO and RHESSI satellites due to their night time or passing the radiation belts. In this regard we underlined three results [11-13] obtained by using ACS SPI data. In the solar event on October 26, 2003 the HXR burst after > 90 min since the impulsive phase was discovered by the ACS SPI data [11], which was not accompanied by any SXR response (burst on tail – BOT). Decay phases of the giant events of the 23rd solar cycle remind those of the June 1991 events with prolonged γ-emission [12]. It was shown in [13] that time profiles of plasma temperature and HXR intensity logarithm were similar in the beginning of the December 6, 2006 event. This correlation disappeared during the decay phase, i.e. significant electron fluxes may not be able to heat effectively the expanding plasma [13].

## 2. Observations

An automated data analysis pipeline was created by the authors of [7] to monitor the high-energy $\gamma$-ray flux from the Sun throughout the Fermi mission. The time intervals in which the Sun is less than $60^0$off axis for the LAT were selected for the analysis [7] and they are called the Fermi Time Windows (FTW). Each interval corresponds to the maximum time with continuous Sun exposure, and the durations of these intervals vary as the Sun advances along the ecliptic and as the orbit of Fermi precesses. The high energy >100 MeV γ-emission was registered by the LAT/FermiGRO instrument in 18 solar flares occurred between June 2010 and July 2012. Since statistics of solar high energy γ-photons is low within FTW's data are averaged for each FTW.

Table 1 summarizes observations by LAT, GBM and ACS SPI for dates in the first column. The second column shows start (UT) and duration (min) of FTW's for LAT solar observations [7]. The third column is a maximal energy of solar emission registered by GBM/FermiGRO. Next columns present characteristics of ACS SPI count rate enhancements: start (UT), duration (min) and maximal value (counts) due to primary HXR; start (UT) due to secondary HXR (SEP arrival). The ACS SPI detector was switched off in 4 from the 18 LAT/FermiGRO events. Bursts of solar HXR >100 keV emission were observed by ACS SPI in the impulsive phase of 13 events. Note that only in six cases FTW's coincide in time with HXR bursts registered by ACS SPI. In these six cases GBM measured HXR emission >300 keV. Time intervals, when prolonged high energy γ-emission was registered by LAT, were not accompanied by HXR >100 keV emission according to the GBM and ACS SPI data (i.e. electrons were accelerated only up to 100 keV). Therefore, most likely, protons producing $\pi^0$ were accelerated during impulsive phase of the flares and then trapped.

**Table 1.** The Fermi LAT high energy γ-events [7] and solar HXR bursts (Fermi GMB http://sprg.ssl.berkeley.edu/~tohban/browser/) and ACS SPI (http://isdc.unige.ch/~savchenk/spiacs-online/). Six cases, when Fermi Time Windows (FTW) and ACS SPI bursts coincide in time, are in bold.

| | FTW | GMB | ACS SPI | | | | |
|---|---|---|---|---|---|---|---|
| | Start,<br>duration | keV | Start,<br>(min) | duration | Max(UT, | counts) | SEP (start, UT) |
| **2010Jun12** | **00:55, 0.8** | **1000** | **00:55:40.8** | **2** | **00:55:50.8** | **4196** | **no** |
| 2011Mar7 | 20:15, 25<br>23:26, 36<br>02:38,35<br>05:49, 35 | <300<br>no<br><50<br><50 | 19:58:00 | 15 | 20:01:54 | 223 | Mar07/ 2143 |
| 2011Jun2 | 09:43, 45 | <50 | No enhancements | | | | |
| 2011Jun7 | 07:34, 53 | <50 | 06:24:00 | 18 | 06:25:49.9 | 107 | Jun 07/ 0646 |
| 2011Aug4 | 04:59, 34 | <25 | 03:50:00 | 9 | 03:51:18.6 | 204 | Aug 04/ 0515 |
| **2011Aug9** | **08:01, 3.3** | **<300** | **08:00:30** | **6.5** | **08:02:10.8** | **732** | **Aug 09/ 0812** |
| **2011Sep6** | **22:17, 0.2**<br>**22:13, 35** | <br>**1000** | **22:18:09** | **6** | **22:19:02.4** | **2075** | **no** |
| 2011Sep7 | 23:36, 63 | <50 | 22:35:51 | 3 | 22:36::18 | 1441 | no |
| **2011Sep24** | **09:34, 0.8** | **1000** | **09:35:00** | **3** | **09:36:32.4** | **5385** | **no** |
| 2012Jan23 | 04:07, 51<br>05:25, 69<br>07:26, 16<br>08:47, 35 | <300<br><25?<br>no<br><50? | No ACS SPI data | | | | |
| 2012Jan27 | 19:45, 11<br>21:13, 24 | <25?<br><25? | 17:38:00 | 9 | 17:38:43 | 185 | Jan 27/ 1755 |
| 2012Mar5 | 04:12, 49<br>05:26, 71<br>07:23, 28 | <300<br><50<br><50 | No ACS SPI data | | | | |
| 2012Mar7 | **00:46, 60**<br><br>03:56, 32<br>07:07, 32<br>10:18, 32<br>13:29, 32<br>19:51, 25 | <300<br>**1000**<br><25<br><25<br><50?<br><50?<br>no | 00:15:00<br>**01:07:00** | >25<br>**30** | 00:17:54<br>**01:07:00** | 1174<br>**1394** | **Mar 07 / 0150** |
| 2012Mar9 | 05:17, 34<br>06:52, 35<br>08:28, 34 | <25<br><100<br>no | No ACS SPI data | | | | |
| 2012Mar10 | 21:05, 30 | <25? | No ACS SPI data | | | | |
| 2012May17 | 02:18, 22 | <100? | 01:29:00 | 10 | 01:35:16 | 85 | May 17/ 0141 |
| **2012Jun3** | **17:40, 23** | **<300** | **17:53:00** | **1.4** | **17:53:24** | **296** | **No** |
| 2012Jul6 | 23:19, 52 | <50 | 23:04:30 | 6.5 | 23:06:34 | 23424 | Jul 06/ 2332 |

**Table 2.** Solar proton events (http://umbra.nascom.nasa.gov/SEP/) and their parent X-ray event.

| | Solar proton event | | X-ray solar event | | | | | |
|---|---|---|---|---|---|---|---|---|
| | Day/UT | $PFU_{max}$<br>/UT | X-class, Coord<br>UT | T(MK)<br>UT | $TTM_T$<br>min | EM<br>$10^{49}cm^{-3}$,<br>UT | $TTM_{EM}$<br>min | Δt<br>min |
| 2010Jun12 | No | | M2.0<br>00:30-01:02 | 20.5<br>00:56 | 26 | 1.0<br>00:58 | 28 | 2 |
| **2011Mar7** | **08/01:05** | **50**<br>**08/08:00** | **M3.7 S22W67**<br>**19:43-20:58** | **13.4**<br>**20:04** | **21** | **2.2**<br>**20:13** | **30** | **9** |
| 2011Jun2 | no | | C2.7<br>09:42-09:50 | 9.7<br>09:47 | 5 | 0.27<br>09:48 | 6 | 1 |
| 2011Jun7 | 07/08:20 | 72<br>07/18:20 | M2.5<br>S21W64<br>06:16-06:59 | 13.6<br>06:28 | 12 | 1.53<br>06:33 | 17 | 5 |
| 2011Aug4 | 04/06:35 | 96<br>05/21:50 | M9.3<br>N19W36<br>03:41-04:04 | 18.0<br>03:54 | 13 | 4.9<br>03:57 | 16 | 3 |
| 2011Aug9 | 09/08:45 | 26<br>09/12:10 | X6.9 N17W69<br>07:48-08:08 | 32.5<br>08:03 | 15 | 29<br>08:05 | 17 | 2 |
| 2011Sep6 | no | | X2.1 N14W18<br>22:12-22:24 | 28.5<br>22:19 | 7 | 9.9<br>22:21 | 9 | 2 |
| 2011Sep7 | no | | X1.8 N14W28<br>22:32-22:44 | 21.8<br>22:37 | 5 | 7.8<br>22:39 | 7 | 2 |
| 2011Sep24 | 23/22:55 | 35<br>26/11:55 | X1.9 N12E60<br>09:21-09:48 | 29.5<br>09:37 | 16 | 7.7<br>09:42 | 21 | 5 |
| **2012Jan23** | **23/05:30** | **6310**<br>**24/15:30** | **M8.7 N28W21**<br>**03:38-04:34** | **17.8**<br>**03:50** | **12** | **4.7**<br>**04:01** | **22** | **10** |
| **2012Jan27** | **27/19:05** | **796**<br>**28/02:05** | **X1.7**<br>**N27W71**<br>**17:37-18:56** | **17.6**<br>**18:27** | **50** | **9.7**<br>**18:40** | **63** | **13** |
| **2012Mar5** | **STEREO B** | | **X1.1 N16E54**<br>**02:30-04:43**<br>**N17E52** | **18.4**<br>**03:37**<br>**19.0**<br>**03:52** | **67**<br><br>**82** | **5.7**<br>**04:11** | **101** | **34**<br><br>**19** |
| **2012Mar7** | **07/05:10** | **6530**<br>**08/11:15** | **X5.4 N17E27**<br>**00:02-00:40**<br>X1.3 N22E12<br>01:05–1:23 | **26.0**<br>**00:18**<br>16<br>01:15 | **16**<br><br>10 | **23.5**<br>**00:25**<br>7.0<br>01:16 | **23**<br><br>10 | **7**<br><br>1 |
| 2012Mar9 | Previous SEP event | | M6.3 N15W01<br>03:22-04:18 | 18.6<br>03:26<br>18.0<br>03:44 | 4<br><br>22 | 0.9<br>03:27<br>3.3<br>03:53 | 5<br><br>31 | 1<br><br>9 |
| 2012Mar10 | Previous SEP event | | M8.4<br>17:15-18:30 | 19.8<br>17:24<br>18.2<br>17:38 | 9<br><br>23 | 2.8<br>17:27<br>4.3<br>17:46 | 12<br><br>31 | 3<br><br>8 |
| **2012May17** | **17/02:10** | **255**<br>**17/04:30** | **M5.1 N11W76**<br>**01:25-02:14** | **15.5**<br>**01:38** | **13** | **2.9**<br>**01:48** | **23** | **10** |
| 2012Jun3 | no | | M3.3<br>17:48-17:57 | 14.9<br>17:54 | 6 | 1.9<br>17:56 | 7 | 1 |
| 2012Jul6 | 07/04:00 | 25<br>07/07:45 | X1.1 S18W50<br>23:01-23:49 | 25.5<br>23:07 | 6 | 5.5<br>23:09 | 8 | 2 |

Table 2 presents information on related proton events (start (UT), maximum (UT) and maximal proton flux). Proton fluxes are integral 5-minute averages for energies >10 MeV, given in Proton Flux Units (PFU), measured by GOES spacecraft 1 PFU = 1 p $cm^{-2}$ $sr^{-1}$ $s^{-1}$. The start of proton events is defined as the first of 3 consecutive data points with fluxes greater than or equal to 10 PFU. Next columns present plasma characteristics of parent solar flares deduced from GOES SXR observations (maximal temperature (MK) and its time moment (UT); maximal emission measure and its time moment; time to maximum (TTM) for temperature and emission measure since the beginning of SXR event, a time delay (min) between their values. SEP events were not registered in 5 from the 18 events even considering additionally observations aboard the STEREO A/B spacecraft [14]. Therefore accelerated protons are not always released into the heliosphere. We note that a difference between time to maximum (TTM) of plasma temperature and emission measure is greater for flares with prolonged high energy γ-emission, which are parent flares of powerful proton events (table 2). These cases are in bold in table 2. At present we do not know a physical reason for such delay. Below three SEP events on March 5, March 7 and July 6, 2012 are considered in more details.

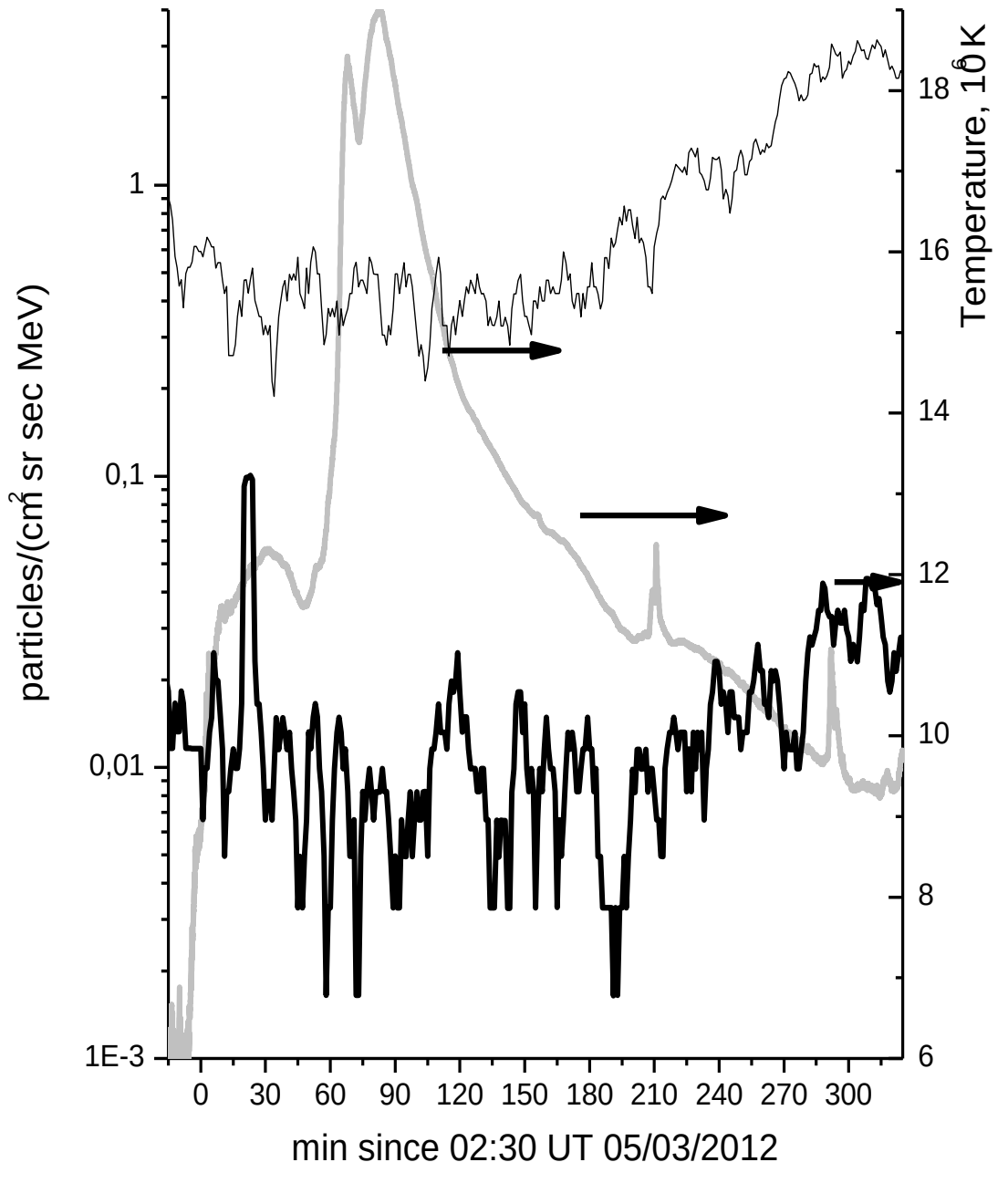

**Figure 1.** Onset of SEP event (thin black line – electrons of 1.4-2.8 MeV and - thick black line protons of 40-60 MeV) observed by HET/STEREO B on March 5, 2012 in comparison with plasma temperature (grey line). Arrows mark FTW's.

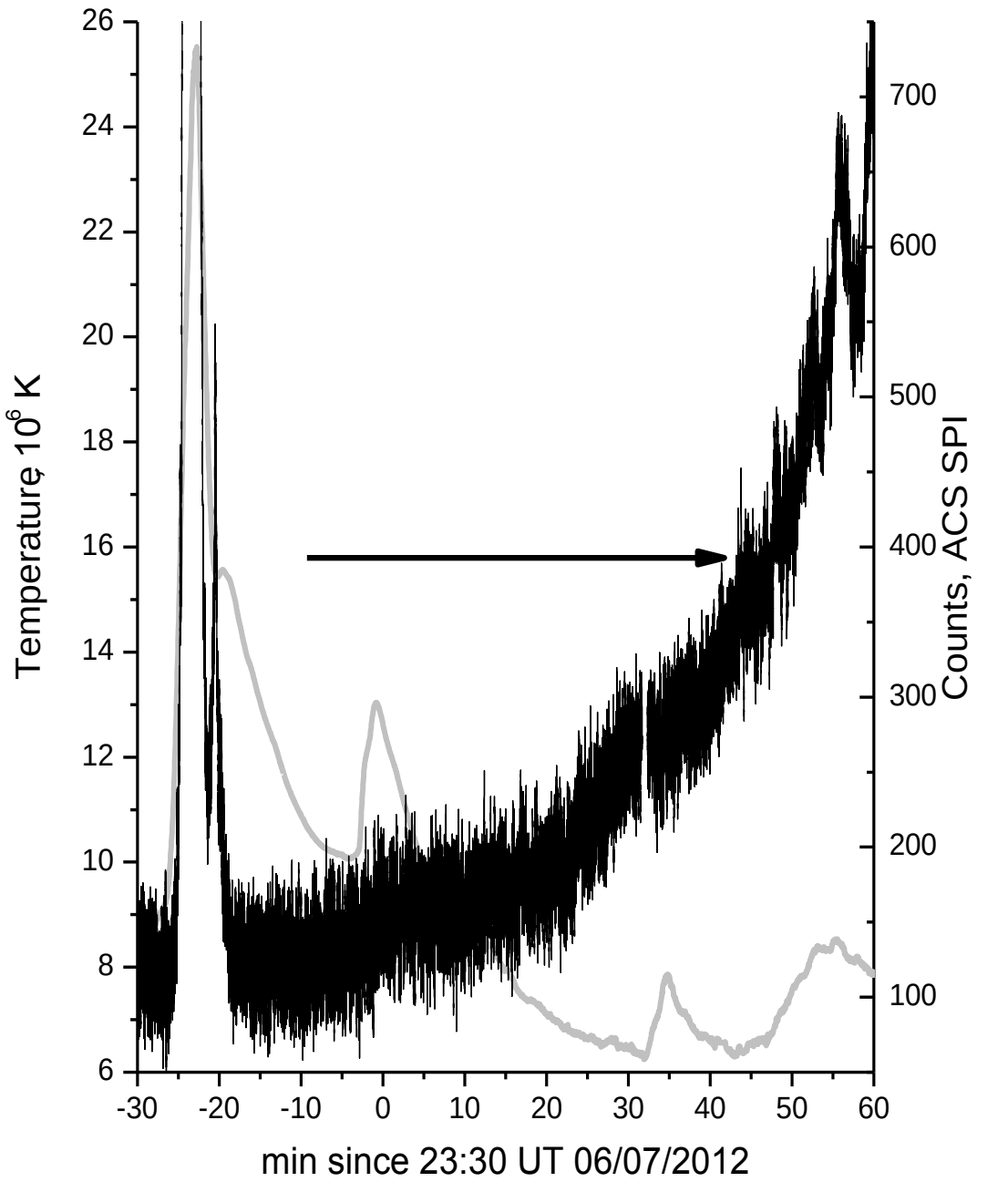

**Figure 2.** Time profiles of plasma temperature (grey line) and ACS SPI count rate (black) Maximal ACS SPI counts of ~23434 are out of scale for the X1.1 event on July 6, 2012. An arrow marks the FTW (52 min from 23:19 UT).

High energy γ-emission was registered on March 5, 2012 within three FTW's (fig. 1). The first FTW corresponds to the flare decay phase, when HXR emission <300 keV was observed by GMB. Maxima of the plasma temperature deduced from GOES SXR observations correspond to the second and third time FTW's, during which only HXR emission <50 keV was registered by GBM. The high energy γ-photons were emitted during periods of energy release associated with electron acceleration to <50 keV. Data from ACS SPI are not available for this event. The SEP event on March 5, 2012 was observed aboard STEREO B, which was in better position for the parent flare with coordinates N16E54 in comparison with SOHO and STEREO-B, at a background of the event on March 4 [14]. The solar electron and proton intensity started to rise during the second FTW (fig. 1). Solar electrons and protons might be released into the heliosphere during time intervals of the Fermi windows.

The event on July 6, 2012 is outstanding, because the maximal count rate of ACS SPI in its impulsive phase was comparable with maximal values observed in the giant events of October-November 2003 [10]. The enhancement of ACS SPI count rate caused by SEP particles possibly masks delayed bursts of primary solar HXR during periods of high energy γ-ray registration as shown in Fig. 2 for the event on July 6, 2012. The GBM didn't show HXR bursts >50 keV during the decay phase. Note that SEP (secondary γ-rays) were not registered by ACS SPI in 5 events, however even in these cases HXR bursts were not observed during the late decay phase.

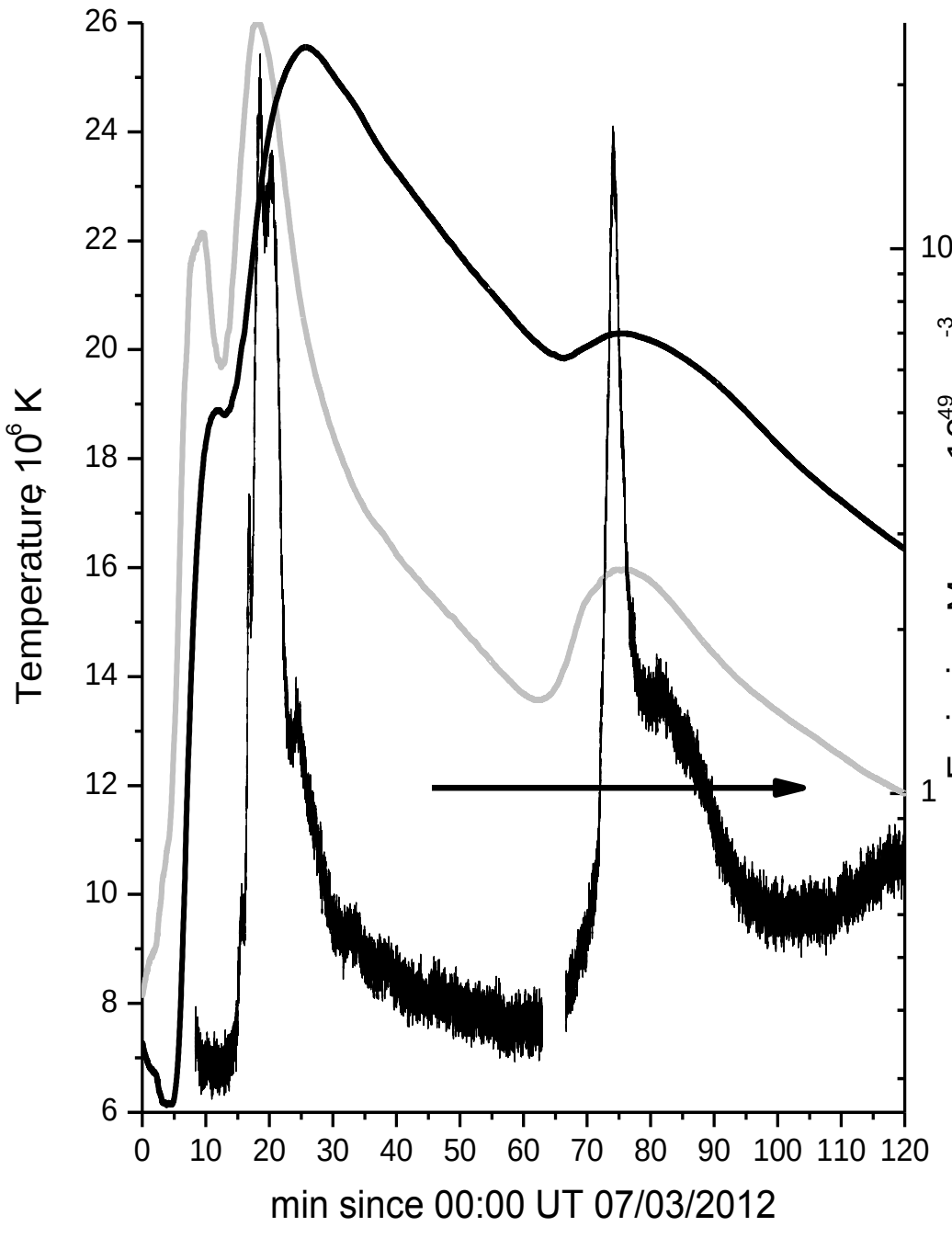

**Figure 3.** Time profiles of plasma temperature (grey line) and emission measure (black line) in comparison with ACS SPI count rate for the X5.4 and X1.3 events on March 7, 2012. An arrow marks the first FTW (60 min from 00:46 UT).

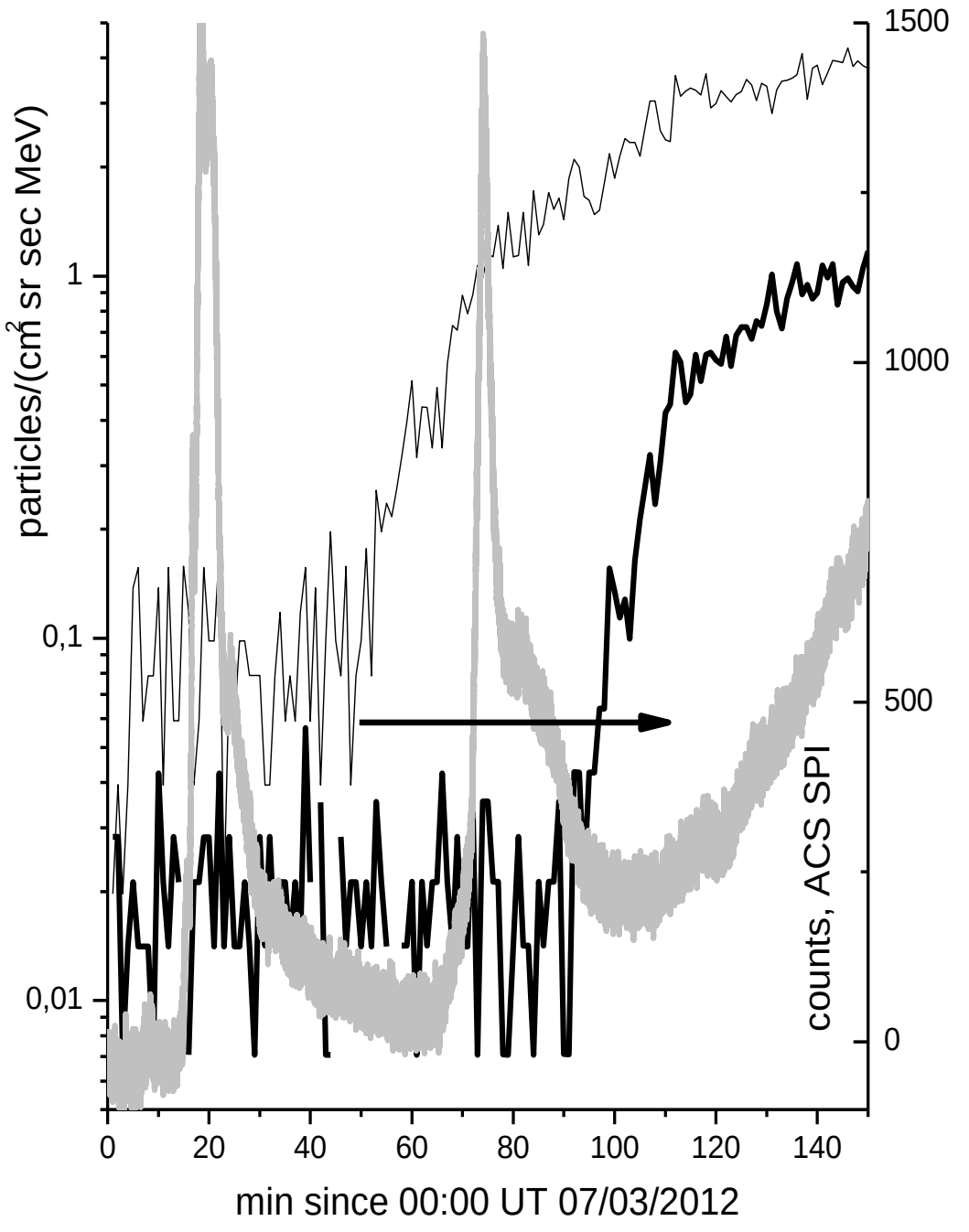

**Figure 4.** Onset of SEP event (thin black line – electrons of 1.4-2.8 MeV and thick black line - protons of 29.5-33.4 MeV) observed by HET/STEREO B on March 7, 2012 in comparison with ACS SPI count rate.

Two different solar flares (X5.4 onset at 00:02 UT and X1.3 onset at 01:05 UT) accompanied by comparable bursts of HXR with time lag of ~60 min composed the event on March 7, 2012 (Fig. 3). By a ratio of effects observed in SXR and HXR in the first and second flares this event reminds the event on October 26, 2003 [9]. The first flare corresponds to impulsive phase, but the second one to BOT. The FermiGRO spacecraft was at the night side of the orbit during the first X5.4 flare. The high energy γ-emission was registered during the second X1.3 flare and later during next five FTW's in the prolonged decay phase [7, 8]. The GBM didn't observe HXR bursts >50 keV within these five FTW's (i.e. electrons were not accelerated to energy >50 keV). Most likely, protons producing $\pi^0$ were accelerated during impulsive phases of these two flares and then trapped.

The authors [14] believe that the well-connected STEREO B observations provide unambiguous evidence of the correct solar association, according to their opinion onset of particle acceleration was associated with the first flare and not with the second one. Their arguments are: 1) there was no additional feature in the intensity–time profile that might be related to the second event; 2) only the first event was accompanied by bright type III radio emissions; 3) interplanetary type II emission commenced at 01:00 UT ahead of the second event. These arguments correspond to our tentative conclusion that flares associated with proton acceleration are characterized by large time delay between maximal temperature and maximal emission measure (table 2). Undoubtedly first electrons registered by STEREO B are associated with the first solar flare as seen in fig. 4. However we may not exclude that protons accelerated in both flares continuously interacted in the solar atmosphere producing $\pi^0$ and released into the heliosphere for more than 20 hours.

## 3. Conclusions

- We do not find any difference between solar flares accompanied by high energy γ-emission and flares without it in other wave bands.
- High energy solar γ-emission observed by LAT/FermiGRO during impulsive phase of flares was accompanied by HXR emission >100 keV observed by GBM/FermiGRO and/or ACS SPI, but not so during prolonged decay phase, therefore processes of proton and electron acceleration should be different in impulsive and prolonged decay phases and/or proton trapping is more effective than electrons.
- Not all flares with high energy γ-emission are associated with SEP events even considering SEP observations at three far-spaced spacecraft in the heliosphere (the STEREO A/B and SOHO spacecraft). If particles interacting in the solar atmosphere and escaping into the heliosphere are of the same population, then processes of particle release into the heliosphere play a crucial role.
- Prolonged observations of high energy gamma and HXR emission with better spatial, temporal and energetic resolution are necessary for progress in the problem of SEP origin.

## Acknowledgment

This work was partly supported by the Presidium of the Russian Academy of Sciences (Program P-22) and the joint Russian-Chinese grant (RFBR-13-02-91165 and NNSFC-11233008). The authors thank two reviewers for their valuable comments and suggestions.